\documentstyle[twoside]{article}

\catcode`\@=11
\long\def\@makefntext#1{
\protect\noindent \hbox to 3.2pt {\hskip-.9pt  
$^{{\eightrm\@thefnmark}}$\hfil}#1\hfill}               

\def\@makefnmark{\hbox to 0pt{$^{\@thefnmark}$\hss}}    
        
\def\ps@myheadings{\let\@mkboth\@gobbletwo
\def\@oddhead{\hbox{}
\rightmark\hfil\eightrm\thepage}   
\def\@oddfoot{}\def\@evenhead{\eightrm\thepage\hfil
\leftmark\hbox{}}\def\@evenfoot{}
\def\sectionmark##1{}\def\subsectionmark##1{}}

\oddsidemargin=\evensidemargin
\newcounter{sectionc}\newcounter{subsectionc}\newcounter{subsubsectionc}
\renewcommand{\section}[1] {\vspace{12pt}\addtocounter{sectionc}{1} 
\setcounter{subsectionc}{0}\setcounter{subsubsectionc}{0}\noindent 
        {\tenbf\thesectionc. #1}\par\vspace{5pt}}
\renewcommand{\subsection}[1] {\vspace{12pt}\addtocounter{subsectionc}{1} 
        \setcounter{subsubsectionc}{0}\noindent 
        {\bf\thesectionc.\thesubsectionc. {\kern1pt \bfit #1}}\par\vspace{5pt}}
\renewcommand{\subsubsection}[1] {\vspace{12pt}\addtocounter{subsubsectionc}{1}
        \noindent{\tenrm\thesectionc.\thesubsectionc.\thesubsubsectionc.
        {\kern1pt \tenit #1}}\par\vspace{5pt}}
\newcommand{\nonumsection}[1] {\vspace{12pt}\noindent{\tenbf #1}
        \par\vspace{5pt}}

\newcounter{appendixc}
\newcounter{subappendixc}[appendixc]
\newcounter{subsubappendixc}[subappendixc]
\renewcommand{\thesubappendixc}{\Alph{appendixc}.\arabic{subappendixc}}
\renewcommand{\thesubsubappendixc}
        {\Alph{appendixc}.\arabic{subappendixc}.\arabic{subsubappendixc}}

\renewcommand{\appendix}[1] {\vspace{12pt}
        \refstepcounter{appendixc}
        \setcounter{figure}{0}
        \setcounter{table}{0}
        \setcounter{lemma}{0}
        \setcounter{theorem}{0}
        \setcounter{corollary}{0}
        \setcounter{definition}{0}
        \setcounter{equation}{0}
        \renewcommand{\thefigure}{\Alph{appendixc}.\arabic{figure}}
        \renewcommand{\thetable}{\Alph{appendixc}.\arabic{table}}
        \renewcommand{\theappendixc}{\Alph{appendixc}}
        \renewcommand{\thelemma}{\Alph{appendixc}.\arabic{lemma}}
        \renewcommand{\thetheorem}{\Alph{appendixc}.\arabic{theorem}}
        \renewcommand{\thedefinition}{\Alph{appendixc}.\arabic{definition}}
        \renewcommand{\thecorollary}{\Alph{appendixc}.\arabic{corollary}}
        \renewcommand{\theequation}{\Alph{appendixc}.\arabic{equation}}
        \noindent{\tenbf Appendix \theappendixc #1}\par\vspace{5pt}}
\newcommand{\subappendix}[1] {\vspace{12pt}
        \refstepcounter{subappendixc}
        \noindent{\bf Appendix \thesubappendixc. {\kern1pt \bfit #1}}
        \par\vspace{5pt}}
\newcommand{\subsubappendix}[1] {\vspace{12pt}
        \refstepcounter{subsubappendixc}
        \noindent{\rm Appendix \thesubsubappendixc. {\kern1pt \tenit #1}}
        \par\vspace{5pt}}

\newcommand{\textlineskip}{\baselineskip=13pt}
\newcommand{\smalllineskip}{\baselineskip=10pt}

\def\eightcirc{
\begin{picture}(0,0)
\put(4.4,1.8){\circle{6.5}}
\end{picture}}
\def\eightcopyright{\eightcirc\kern2.7pt\hbox{\eightrm c}} 

\newcommand{\copyrightheading}[1]
        {\vspace*{-2.5cm}\smalllineskip{\flushleft
        {\footnotesize International Journal of Modern Physics A, #1}\\
        {\footnotesize $\eightcopyright$\, World Scientific Publishing
         Company}\\
         }}

\def\abstracts#1#2#3{{
        \centering{\begin{minipage}{4.5in}\baselineskip=10pt\footnotesize
        \parindent=0pt #1\par 
        \parindent=15pt #2\par
        \parindent=15pt #3
        \end{minipage}}\par}} 

\newcommand{\bibit}{\nineit}

\renewenvironment{thebibliography}[1]
        {\frenchspacing
         \ninerm\baselineskip=11pt
         \begin{list}{\arabic{enumi}.}
        {\usecounter{enumi}\setlength{\parsep}{0pt}
         \setlength{\leftmargin 12.7pt}{\rightmargin 0pt} 
         \setlength{\itemsep}{0pt} \settowidth
        {\labelwidth}{#1.}\sloppy}}{\end{list}}

\newcounter{itemlistc}
\newcounter{romanlistc}
\newcounter{alphlistc}
\newcounter{arabiclistc}

\newcommand{\fcaption}[1]{
        \refstepcounter{figure}
        \setbox\@tempboxa = \hbox{\footnotesize Fig.~\thefigure. #1}
        \ifdim \wd\@tempboxa > 5in
           {\begin{center}
        \parbox{5in}{\footnotesize\smalllineskip Fig.~\thefigure. #1}
            \end{center}}
        \else
             {\begin{center}
             {\footnotesize Fig.~\thefigure. #1}
              \end{center}}
        \fi}

\newcommand{\tcaption}[1]{
        \refstepcounter{table}
        \setbox\@tempboxa = \hbox{\footnotesize Table~\thetable. #1}
        \ifdim \wd\@tempboxa > 5in
           {\begin{center}
        \parbox{5in}{\footnotesize\smalllineskip Table~\thetable. #1}
            \end{center}}
        \else
             {\begin{center}
             {\footnotesize Table~\thetable. #1}
              \end{center}}
        \fi}

\def\@citex[#1]#2{\if@filesw\immediate\write\@auxout
        {\string\citation{#2}}\fi
\def\@citea{}\@cite{\@for\@citeb:=#2\do
        {\@citea\def\@citea{,}\@ifundefined
        {b@\@citeb}{{\bf ?}\@warning
        {Citation `\@citeb' on page \thepage \space undefined}}
        {\csname b@\@citeb\endcsname}}}{#1}}

\newif\if@cghi
\def\cite{\@cghitrue\@ifnextchar [{\@tempswatrue
        \@citex}{\@tempswafalse\@citex[]}}
\def\citelow{\@cghifalse\@ifnextchar [{\@tempswatrue
        \@citex}{\@tempswafalse\@citex[]}}
\def\@cite#1#2{{$\null^{#1}$\if@tempswa\typeout
        {IJCGA warning: optional citation argument 
        ignored: `#2'} \fi}}

\def\pmb#1{\setbox0=\hbox{#1}
        \kern-.025em\copy0\kern-\wd0
        \kern.05em\copy0\kern-\wd0
        \kern-.025em\raise.0433em\box0}

\def\fnt#1#2{\footnotetext{\kern-.3em
        {$^{\mbox{\scriptsize #1}}$}{#2}}}

\def\fpage#1{\begingroup
\voffset=.3in
\thispagestyle{empty}\begin{table}[b]\centerline{\footnotesize #1}
        \end{table}\endgroup}

\def\runninghead#1#2{\pagestyle{myheadings}
\markboth{{\protect\footnotesize\it{\quad #1}}\hfill}
{\hfill{\protect\footnotesize\it{#2\quad}}}}
\font\tenrm=cmr10
\font\tenit=cmti10 
\font\tenbf=cmbx10
\font\bfit=cmbxti10 at 10pt
\font\ninerm=cmr9
\font\nineit=cmti9

\font\eightrm=cmr8

\def\qed{\hbox{${\vcenter{\vbox{                        
   \hrule height 0.4pt\hbox{\vrule width 0.4pt height 6pt
   \kern5pt\vrule width 0.4pt}\hrule height 0.4pt}}}$}}

\begin{document}

\runninghead{Laboratory Limits on Theories with Sterile Neutrinos in
  the Bulk} {Ara Ioannisian and Apostolos Pilaftsis}

\normalsize\textlineskip
\thispagestyle{empty}
\setcounter{page}{1}

\copyrightheading{}                     
\vspace{-0.8cm}
\begin{flushright}
WUE-ITP-2000-029\\
hep-ph/0010051
\end{flushright}

\vspace*{0.7truein}
\fpage{1}
\centerline{\bf LABORATORY LIMITS ON THEORIES WITH}
\vspace*{0.035truein}
\centerline{\bf STERILE NEUTRINOS IN THE BULK }
\vspace*{0.37truein}
\centerline{\footnotesize ARA IOANNISIAN\footnote[1]{On leave of absence
from Yerevan Physics Institute, Alikhanian Br.\ 2, 375036 Yerevan,
Armenia}}
\vspace*{0.015truein}
\centerline{\footnotesize\it Instituto de F\'{\i}sica Corpuscular --
C.S.I.C. -- Universitat of Val\`encia}
\baselineskip=10pt
\centerline{\footnotesize\it Edificio Institutos de Paterna -- Apartado de
Correos 2085 - 46071  Val\`encia, Spain
} \vspace*{10pt} 
\centerline{\footnotesize APOSTOLOS PILAFTSIS\footnote[2]{Talk given
  at the DPF2000 (August 9-12, 2000), hosted by  the Ohio 
  State University, Columbus, Ohio, USA.}
}
\vspace*{0.015truein} \centerline{\footnotesize\it Institut f\"ur
  Theoretische Physik, Universit\"at W\"urzburg, } \baselineskip=10pt
\centerline{\footnotesize\it Am Hubland, 97074 W\"urzburg, Germany}

\vspace*{0.21truein}  \abstracts{ We   discuss the    phenomenological
  consequences of theories  which describe sterile neutrinos in  large
  extra dimensions, in the so-called bulk.  We briefly outline how the
  cumulative non-decoupling effect  due to  the tower of  Kaluza-Klein
  singlet  neutrinos    may    equivalently    be described     by   a
  higher-dimensional effective theory with original order-unity Yukawa
  couplings.  Based on  this cumulative phenomenon, we obtain strong
  constraints  on the fundamental  quantum gravity scale and/or on the
  higher-dimensional Yukawa couplings.}{}{}

\textlineskip                   
\vspace*{12pt}                  

\vspace*{-0.5pt}
\noindent

It has recently been realized \cite{ADD} that the experienced weakness
of   gravity may be  ascribed to   the  spreading of the gravitational
forces in  $\delta\ (\ge 2)$  large extra (spatial) dimensions, namely
in the so-called  bulk.  In  such a  novel scenario,  the  fundamental
$\sim$~TeV  scale of quantum gravity  $M_F$ is related to the ordinary
Planck     mass  $M_{\rm   P}    =  1.2\times   10^{19}$~GeV and   the
compactification scale $R$, through  a kind of generalized Gauss' law:
$M_{\rm P}   \approx M_F (RM_F)^{\delta/2}$.   In  this framework, the
Standard  Model (SM) fields  are confined to  a 3-brane configuration,
whilst  gravity and most probably  fields which are singlets under the
SM gauge  group,   such as sterile   neutrinos,  are only allowed   to
propagate in the large compactified dimensions.\cite{ADDM,DDG}

For  our  phenomenological  discussion,   we  shall  adopt  a  variant
\cite{AP,IN} of the  higher-dimensional  singlet-neutrino model  which
was originally discussed  in Refs.\cite{ADDM,DDG} For definiteness, we
consider a  model that minimally extends the  SM-field  content by one
singlet  Dirac   neutrino,     $N(x,y)$,   which  propagates    in   a
$[1+(3+\delta)]$-dimensional  Minkowski space. The $y$-coordinates are
compactified   on a circle  of  radius  $R$  by  applying the periodic
identification: $y \equiv y + 2\pi R$.  For more details on this model
as well as  the conventions followed here, the  reader  is referred to
Ref.\cite{IN}

The leptonic  sector of the  our minimal  model  of interest to  us is
given by
\begin{equation}
  \label{Leff}
{\cal L}_{\rm eff} = \int\limits_0^{2\pi R}\!\! d\vec{y}\
 \Big[\, \bar{N} \Big( i\gamma^\mu \partial_\mu\, +\,
 i\gamma_{\vec{y}} \partial_{\vec{y}} -m \Big) N\ +\ \delta (\vec{y})\,
\Big( \sum_{l=e,\mu,\tau}\, \frac{h_l}{M_F^{\delta/2}} L_l\tilde{\Phi}
\xi\, +\, {\rm h.c.}\,\Big) \,\Big],
\end{equation}
where  $\tilde{\Phi}  = i\sigma_2 \Phi$   ($\langle\tilde{\Phi}\rangle
=v=174$~GeV)  and    $L_l$ are  the     Higgs and   lepton   doublets,
respectively.  In  Eq.~(\ref{Leff}),   we have assumed that   only one
two-component spinor $\xi$ of the higher-dimensional one $N$ interacts
with our  3   brane, while   its  non-interacting two-component  Dirac
partner   is   denoted by    $\eta$.    After   integrating out    the
$y$-coordinates, we arrive at the effective Lagrangian:
\begin{equation}
  \label{LKK} 
\sum_{\vec{n}}
\bigg[ \bar{\xi}_{\vec{n}}
( i\bar{\sigma}^\mu \partial_\mu) \xi_{\vec{n}} +
\bar{\eta}_{\vec{n}} ( i\bar{\sigma}^\mu \partial_\mu) \eta_{\vec{n}}
- \Big( \xi_{\vec{n}}(m + \frac{i|\vec{n}|}{R}) \eta_{-{\vec{n}}}
+\sum_{l=e,\mu,\tau}
\bar{h}_l L_l\tilde{\Phi} \xi_{\vec{n}} + {\rm h.c.} \Big) \bigg]
\end{equation}
where $|\vec{n}|^2\equiv \sum_{i=1}^{\delta}  n_i^2$,  $\xi_{\vec{n}}$
and   $\eta_{\vec{n}}$  are  the Kaluza-Klein   (KK)  excitations, and
$\bar{h}_l   =   h_l  M_F/M_{\rm  P}$   is   the   known 4-dimensional
Yukawa-coupling suppressed  by the  volume  factor $M_F/M_{\rm  P}\sim
10^{-16}$.\cite{ADDM,DDG} In this  minimal  model, the observed  light
neutrinos $\nu_l$ have  very small admixtures $B_{l,\vec{n}}$ with the
heavy KK states $\eta_{\vec{n}}$:
\begin{equation}
\label{nonuni}
\nu_l \approx \frac{1}{\sqrt{1+\sum_{\vec{n}}|B_{l,\vec{n}}|^2}}
( \nu_{lL} + \sum_{\vec{n}} B_{l,\vec{n}}\eta_{\vec{n}} )\,,\quad
{\rm with}\quad B_{l,\vec{n}} \approx 
\frac{h_l\,v M_F}{M_P\sqrt{m^2+\vec{n}^2/R^2}}\ ,
\end{equation}
where $\sqrt{m^2+\vec{n}^2/R^2}$ are approximately  the masses  of the
physical heavy KK states. 

In order to quantify the  new-physics effects mediated by KK neutrinos
both at the  tree and one-loop levels, it  proves useful to define the
mixing parameters
\begin{eqnarray}
  \label{snul} (s^{\nu_l}_L)^2 \equiv \sum_{\vec{n}} |B_{l,\vec{n}}|^2
\approx h^2_l\frac{v^2}{M_F^2}\sum_{\vec{n}} \frac{M_F^4
M_P^{-2}}{(m^2+\frac{\vec{n}^2}{R^2})}\approx \left\{ \begin{array}{l}
\frac{\displaystyle \pi h^2_l v^2}{\displaystyle M^2_F}\,
\ln\bigg(\displaystyle \frac{M^2_F}{m^2}+1\bigg),\
\delta=2 \\ \\ \frac{\displaystyle S_\delta}{\displaystyle \delta - 2}\,
\frac{\displaystyle h^2_l v^2}{\displaystyle M^2_F},\ \delta > 2\, ,  
\end{array}\ \right.  
\end{eqnarray} 
where $S_\delta = 2\pi^{\delta/2}/\Gamma  (\delta /2)$ is  the surface
area of a $\delta$-dimensional sphere of unit radius.  In deriving the
last step in  Eq.~(\ref{snul}), we have approximated  the sum over the
KK states by  an integral, with an upper  ultra-violet (UV)  cutoff at
$M_F  R$, above which  string-threshold  effects  are expected to   be
more relevant.

As  can be seen from  Table \ref{Tab1},  the mixings $(s^{\nu_l}_L)^2$
may be constrained by a  number of new-physics observables induced  at
the tree  level.  These observables measure  possible non-universality
effects in $\mu$, $\tau$ and $\pi$ decays.   In this respect, in Table
\ref{Tab1} we have defined  $R_\pi =\Gamma(\pi \to e\nu)/\Gamma(\pi\to
\mu\nu)$  and  $R_{\tau \mu}=B(\tau  \to e   \nu \nu)/B(\tau  \to  \mu
\nu\nu)$.

\begin{table}[t]
\tcaption{Tree-level limits on $M_F/h$.}
\label{Tab1}
\begin{center}
\begin{tabular}{c|cc|cc}
\hline
 & $h_e\ =\ h_\mu$ &\hspace{-1cm}$=\ h_\tau\ =\ h$ &
\hspace{1.cm}$h_\mu =0$ &\hspace{-0.5cm}~and $h_e = h_\tau$ \\
Observable & Lower ~limit &
\hspace{-0.6cm}on $M_F/h$ & Lower ~limit
&\hspace{-0.6cm}on $M_F/h_\tau$ \\
& $\delta =2$ & $\delta > 2$ & $\delta =2 $ & $\delta > 2$ \\
\hline
$1 - \frac{\displaystyle \Gamma(\mu\to e\nu\nu )}{\displaystyle
\Gamma_{\rm SM} (\mu\to e\nu\nu )}$
 & $8.9\,\ln^{1/2}\frac{\displaystyle M_F}{\displaystyle m}$ &
$\frac{\displaystyle 3.5\, S^{1/2}_\delta}{\displaystyle \sqrt{\delta -2}
}$&
$6.3\,\ln^{1/2}\frac{\displaystyle M_F}{\displaystyle m}$&
$\frac{\displaystyle 2.5\, S^{1/2}_\delta}{\displaystyle \sqrt{\delta -2}
}$\\
\hline
$1 - \frac{\displaystyle \Gamma(Z\to \nu\nu )}{\displaystyle
\Gamma_{\rm SM} ( Z\to \nu\nu )}$
 & $5.9\,\ln^{1/2}\frac{\displaystyle M_F}{\displaystyle m}$ &
$\frac{\displaystyle 2.4\, S^{1/2}_\delta}{\displaystyle \sqrt{\delta
-2}}$&
$4.8\,\ln^{1/2}\frac{\displaystyle M_F}{\displaystyle m}$&
$\frac{\displaystyle 1.9\, S^{1/2}_\delta}{\displaystyle \sqrt{\delta
-2}}$\\
\hline
$1 - \frac{\displaystyle R_\pi }{\displaystyle R^{\rm SM}_\pi }$
        & $-$ & $-$ &
$18.7\,\ln^{1/2}\frac{\displaystyle M_F}{\displaystyle m}$&
$\frac{\displaystyle 7.5\, S^{1/2}_\delta}{\displaystyle \sqrt{\delta
-2}}$\\
\hline
$1 - \frac{\displaystyle R_{\tau \mu}}{\displaystyle R^{\rm SM}_{\tau \mu}
}$
 & $-$ & $-$ & $5.7\,\ln^{1/2}\frac{\displaystyle M_F}{\displaystyle m}$&
$\frac{\displaystyle 2.3\, S^{1/2}_\delta}{\displaystyle \sqrt{\delta
-2}}$\\
\hline
\end{tabular}
\end{center}
\end{table}
Owing to the tower of the KK singlet neutrinos which acts cumulatively
in   the loops,     significant  universality-breaking  as   well   as
flavour-violating  effects   are  induced  in   electroweak  processes
involving  $\gamma$-\cite{FP} and $Z$-  boson~\cite{IN}  interactions. 
In particular, as has been explicitly shown recently,\cite{IN} we find
that the  cumulative presence of the KK  states leads to  an effective
theory whose  Yukawa interactions are  mediated  by order-unity Yukawa
couplings of the  original Lagrangian before compactification. In this
case,  we    expect  a higher-dimensional    non-decoupling phenomenon
analogous to the  one studied earlier  in renormalizable 4-dimensional
theories.\cite{IP,PS}      For      example,       the       effective
lepton-flavour-violating vertex $Zll'$   that occurs in $\mu \to  eee$
exhibits  the  dependence:   ${\cal   T}(Zl'l)  \propto h_l   h_{l'}\,
(v^2/M^2_F)\,  \sum_{k=e,\mu,\tau}\,  (h^2_k   v^2)/M^2_W$, i.e.\  its
strength  increases  with the fourth   power of the higher-dimensional
Yukawa  couplings.  This  should   be contrasted  with the  respective
photonic  amplitude   ${\cal  T}(\gamma   l'l) \propto    h_l h_{l'}\,
v^2/M^2_F$, whose strength increases only quadratically.
\begin{table}[t]
\tcaption{One-loop-level limits on $M_F/h^2$.}
\label{Tab2}
\begin{tabular}{cccc}
\hline
 & $h_e\ =\ h_\mu$ &\hspace{-1cm}$=\ h_\tau\ =\ h\ \ge\ 1$ & \\
Observable & Lower ~limit &
\hspace{-0.2cm}on $M_F/h^2\ [\,{\rm TeV}\,]$& \\
& $\delta =2$ & $\delta = 3$ & $\delta =6 $  \\
\hline
Br$(\mu \to e \gamma)$ & $75$ & $43$ & $33$\\
\hline
Br$(\mu \to e e e)   $ & $250$ & $230$ & $200$\\
\hline
Br$(\mu \ ^{48}_{22}{\rm Ti} \to e \ ^{48}_{22}{\rm Ti})$ & $380$ & $320$
&300\\
\hline
\end{tabular}
\end{table}

Based on this cumulative non-decoupling  effect, we are able to derive
strong limits  on  the $M_F/h^2$, for $h\ge   1$.  As is  displayed in
Table~\ref{Tab2}, the   strongest  limits  are   obtained from    $\mu
\not\!\to eee$ and  the absence of  $\mu$-to-$e$ conversion in nuclei. 
Of  course,  the    limits  presented  here  contain    some degree of
uncertainty, which is  inherent  in all effective   non-renormalizable
theories with a   cut-off  scale, such  as  $M_F$.   Nevertheless, our
results  are very  useful, since   they indicate the  generic  size of
constraints that one has to encounter in model-building considerations
with sterile neutrinos.\cite{KKneutr}

\vspace{-0.2cm}
\nonumsection{References}

\end{document}